# Preserving Culinary Traditions. A Crowdsourced Digital Collection of Cookbooks

Giulia Renda[1], Giulia Manganelli[2], Mila Fumini[3], Marilena Daquino[4]

[1] University of Bologna, Italy - giulia.renda3@unibo.it
[2] Independent researcher, Italy - giuliamanganelli@me.com
[3] Independent researcher, Italy - mila.fumini@me.com
[4] University of Bologna, Italy - marilena.daquino2@unibo.it

## ABSTRACT

Recipes of popular origin and handwritten cookbooks are often overlooked by scholars. *Ragù* is a pilot project that tries to fill in this gap by gathering and digitising a collection of cookbooks belonging to the Italian traditional cuisine, and making it accessible via a digital platform. The project aims at contributing to two research lines: a) to identify agile methods for publishing data in a low-cost crowdsourcing project, and b) to devise an effective storytelling journey for the *Ragù* project.

## 1. INTRODUCTION

Recipes of popular origin and handwritten cookbooks are often overlooked by scholars, despite representing a rich source of cultural insight, since they present the influence of diverse cultures and their relationship with traditions, and witness the evolution of material history. Yet, extensive projects to collect, transcribe, and preserve such a precious intangible cultural heritage are missing. *Ragù* is a pilot project that tries to fill in this gap by collecting and digitising a collection of cookbooks belonging to the Italian traditional cuisine, and making it accessible via a digital platform.

Being a low-budget project managed by a few volunteers, the project presents several criticalities. Firstly, the primary sources need to be physically collected, transcribed and digitised and a mid-term data management plan for preservation and exploitation is needed. Additionally, mechanisms for effective metadata extraction and query must be designed to populate user-friendly interfaces. Given the resource constraints of the project, such methods must ensure easy updates (to both metadata and user interfaces) and access to data and images. Lastly, since the web platform aims to engage both everyday users open to serendipitous discovery and scholars interested in exploring data with more sophisticated queries, appropriate strategies for dissemination must be designed.

In this article, we present the strategies adopted in the *Ragù* project. We describe existing approaches to crowdsourcing and storytelling (section 2), we propose a simplified model for data management in low-cost projects (section 3), and we present a digital storytelling journey for the case study at hand (section 4).

## 2. STATE OF THE ART

In the last decades, the massive digitisation process of large libraries, museums and archives has been successfully carried out by specialists in galleries, libraries, archives and museums (GLAM) [9]. However, the business model applied by national and international institutions is not always applicable to small and medium-sized institutions, cultural associations or individuals, who may have to find alternative strategies.

Crowdsourcing [6] is the process of outsourcing tasks or problems to a large group of people, typically using the web as a platform for information exchange. Crowdsourcing methods have been adopted in GLAM and heritage organisations to carry out several tasks, such as metadata creation, analysis of cultural heritage objects, as well as contributions of private objects or experiences [13, 1, 10]. Notably, public collections rarely contain materials belonging to private collections or individuals, e.g. household documents. Therefore crowdsourcing such primary sources may be the only way to foster research enquiries [14] based on witnesses of our material history. An exemplar case is the Europeana 1914-1918[1] campaign, where users could upload pictures of memorabilia to be digitised and populate a digital archive.

However, similar projects on a smaller scale in terms of both finances and workforce encounter problems related to continuity and sustainability [3], including socio-technical, technological, and dissemination challenges. As a matter of fact, most crowdsourcing projects in GLAM rely on expensive and not easy-to-customise platforms for collecting citizens' contributions [3].

---

[1] https://www.europeana.eu/en/collections/topic/83-world-war-i

More importantly, existing projects do not often provide dissemination interfaces to communicate results of the crowdsourcing campaign and may not be able to engage a wide variety of visitors. From the literature on approaches for information seeking purposes, two types of interactive behaviour can be identified. The first strategy focuses on reducing the cognitive load of users by breaking down information, displaying it in small chunks [7]. The second approach advocates for more *generous interfaces*, arguing that hiding the entirety of content from the user can become a source of frustration [16]. Considering that visitors on cultural heritage websites don't always have a specific task in mind, methods for casual browsing can effectively help them to discover and find their way to a specific task or interest. Such an approach has been summarised as "overview first, zoom and filter, then details on demand" [12]. To overcome the overwhelming feeling caused by the large amount of materials that a digital collection can contain, it is common practice in the cultural heritage domain to design "paths" to help the user navigate content, and to use data visualisation techniques to provide overviews and improve their experience [4]. However, data visualisations can be hard to understand for everyday users. To mitigate this complexity, storytelling techniques can be used to convey complex information compactly [5, 8]. The telling of a story (path) acts as an interpretative framework [2].

Examining projects that use storytelling and data visualisation as tools for presenting digital collections of cookbooks and guiding the user exploration, we could find very few examples. *The Early American Cookbooks* [15] collects cookbooks from 1800 to 1920, and serves them on a website organised in thematic categories, which feature blog-like articles using data visualisations to offer insights into the collection. Developed by a digital humanist at the New York University Libraries, the workflow for data management does not seem to be replicable, and the storytelling strategies are limited to few reiterated charts that do not offer means to continue the exploration in depth. Other examples of collections of private cookbooks include *The South African Jewish Cookbook Project*[2], but the project website does not present the collection using data visualisations or storytelling strategies. Rather it offers traditional interfaces for exploring items of the collection thematically. Therefore, this article aims to contribute to two lines of research: a) to identify agile methods for publishing data in a low-cost project (section 3), and b) to devise an effective storytelling journey for crowdsourced cookbooks collections (section 4).

## 3. AN OPEN SOURCE AND REUSABLE APPROACH TO MANAGE LOW-COST CROWDSOURCING CAMPAIGNS

The project began with a serendipitous discovery of handwritten cookbooks in a cellar to be cleared out. Historian Mila Fumini, leading the project, initiated the first call for contributions in 2019. Collection efforts have mainly occurred through public appeals at food-related events or locations, with additional outreach via email, followed by visits to homes to assess potential materials. The only criterion for inclusion in the collection is that notebooks must be handwritten. Whenever a new cookbook is proposed and evaluated, it is photographed with a smartphone and then returned to the owner. Pictures[3] are archived using a private online folder, and selected contents are transcribed in a shared spreadsheet online. The table was devised by a restricted group of data curators that volunteered in the project and it includes bibliographic metadata of cookbooks and recipes - e.g. recipe title, year, city of origin, author - information on the image file, and other insights relevant to historical research enquiries, e.g. type of recipe, ingredients, quantities, and linguistic variations of names (Figure 1). Controlled vocabularies and folksonomies have been defined for ingredients and categories, geographical information (city, region, and country), scope, procedure, and units of measurement. Future plans include aligning with existing resources. The working table is available as a Google Spreadsheet[4] and can be downloaded in CSV format for analysis purposes.

**Figure 1. An extract from the Google Spreadsheet.**

---

[2] https://sajewishcookbooks.org.za/
[3] Available under a CC-BY 4.0 licence.
[4] https://docs.google.com/spreadsheets/d/1terFx_mYVspOjvDUJfcBqHR7j_OlvXVwUesHNHAl_wU/edit?usp=sharing

Digitisation and transcription is an ongoing process that began in 2023, with currently digitised materials spanning approximately 460 recipes from the 1960s to the 1980s. To showcase (partial) results of the data collection as soon as possible, we decided to create a website that could be easily and dynamically updated with new recipes. To achieve this, we leverage one of the most well-known platforms for depositing open source code, i.e. GitHub. GitHub provides the *Ragù* project with four important features, namely: (1) free hosting of data dumps, i.e. a versioned copy of the aforementioned working table, (2) free-of-charge hosting and publication of a static website, (3) editorial strategies to prevent people to immediately publish content on the website before a review of the table has been done, and (4) GitHub Actions to extract data from the versioned table and populate website interfaces dynamically. The GitHub repository of *Ragù* project is available online[5], where documentation is available.

Updating the user interface requires an editor to download the working table as .tsv files (including both the metadata and vocabulary tables), and upload them to the project GitHub repository. Committing the uploaded files is labelled as a "data update", and non-expert users can upload the data using a user-friendly drag-and-drop interface. Images are uploaded similarly in the *recipe_photos* folder. These two types of submissions trigger a GitHub action that runs the *script.py* file, which includes methods to extract information from aforementioned files and write a few configuration files. JSON format was chosen since it is lightweight, it is easily manipulated by both Python and JavaScript languages, it is highly readable also by humans, and in the future it can be easily converted into JSON-LD format to add semantics.

The configuration files automatically created/updated by GitHub actions are used to populate the website thanks to a set of JavaScript scripts. These files are:
- general.json contains statistics to be used in the homepage - e.g. the number of recipe books, recipes, and ingredients - and paths to other files (recipe books and filters).
- Other files contain recipes grouped by letter (alphabet.json), type of course (categories.json), ingredients (ingredients.json) and provenance (provenance.json), which are used to populate the advanced search of the website.
- Bespoke files are created to include data necessary to build visualisations, and are divided by type, namely: map.json, matrix.json, network.json and piechart.json, each organised according to the requirements of Charts.js[6], the data visualisation library chosen to create the charts.
- Finally, each cookbook has its own file, in which we find general information such as title, year, origin, and a dictionary including all the recipes, which in turn contain the annotations extracted from the working table.

Such an approach has several advantages, namely: (1) it leverages the power of the hosting platform, hence preventing us from developing sophisticated custom code for ensuring the continuous update; (2) it prevents non-technical people involved in the project to understand complex concepts related to data management and web development, e.g. it does not require them to learn sophisticated software solutions for crowdsourcing (tables are relatively easy to understand), update (drag-and-drop interfaces hide complexities of version-control-systems), and publication (the web publication is completely automated); (3) it is a free-of-charge, potentially sustainable in the mid-term, solution for hosting and versioning code (at the moment), which prevents us from reserving resources for maintenance of a dedicated server; (4) allows one to perform editorial control without requiring yet another interface to learn (only collaborators of the repository can upload the versioned tables that trigger the update of the website).

## 4. A STORYTELLING PROJECT ON COOKBOOKS

To accommodate requirements of different target audiences, the digital collection is presented on a website[7] that offers three macro-sections, namely: an introductory storytelling journey (*homepage*), a search page based on filters and facets (*recipes*), and an ebook-like browser of digitised cookbooks (*cookbooks*, currently under construction).

The *homepage* offers a digital storytelling journey that exploits the narrative element to showcase the content of the collection and stimulate curiosity. The concept idea leverages the red thread of history and tradition, and uses animation and *scrollytelling*[8] to collate several sections that include charts accompanied by a (sequential) introductory text (Figure 2). This narrative uses data visualisation elements to present collected data in an intuitive way and offers insights that can be leveraged in specific searches later - following the approach of generous interfaces and "overview first, details on demand". In detail, the red string graphically conducts users through five topics that invite them to explore recipes and cookbooks sections, showing significant data patterns that can be appreciated by a broad audience, namely:

1) **Overview**. Users are introduced to the topic (cookbooks), the goal of the project (preservation and dissemination) and are provided with an overview of data, e.g. number of cookbooks, recipes, and ingredients.
2) **Provenance**. The journey continues with a map displaying the provenance of collected cookbooks that stimulates empathy (a user will likely look for recipes from their region) showing where the project originated, and how it evolved engaging with more donors of handwritten notebooks (currently the web application includes only a selection of digitised cookbooks and more have to be included).
3) **Ingredients**. Three data visualisations give the visitor an overview of ingredients: recipes grouped by ingredients show the importance of certain ingredients in the Italian diet, the co-occurrence/correlation of ingredients lets users grasp the constituents of the Italian traditional cuisine, and the analysis of units of measures show them how the experience of the cook has a pivotal role in the Italian intangible heritage.
4) **Dishes**. A set table shows all the types of courses that are collected.
5) **Gender**. A final remark is put on the fact that almost all recipes are written by women.

---

[5] https://github.com/raguproject/raguproject.github.io
[6] https://www.chartjs.org/
[7] https://raguproject.github.io/
[8] The term *scrollytelling* is a combination of "storytelling" and "scrolling". It refers to a technique used in web design and digital journalism where narrative content unfolds as the user scrolls down a web page. *Scrollytelling* allows users to explore content at their own pace, while maintaining their attention and deepening their understanding of the topic being presented [11].

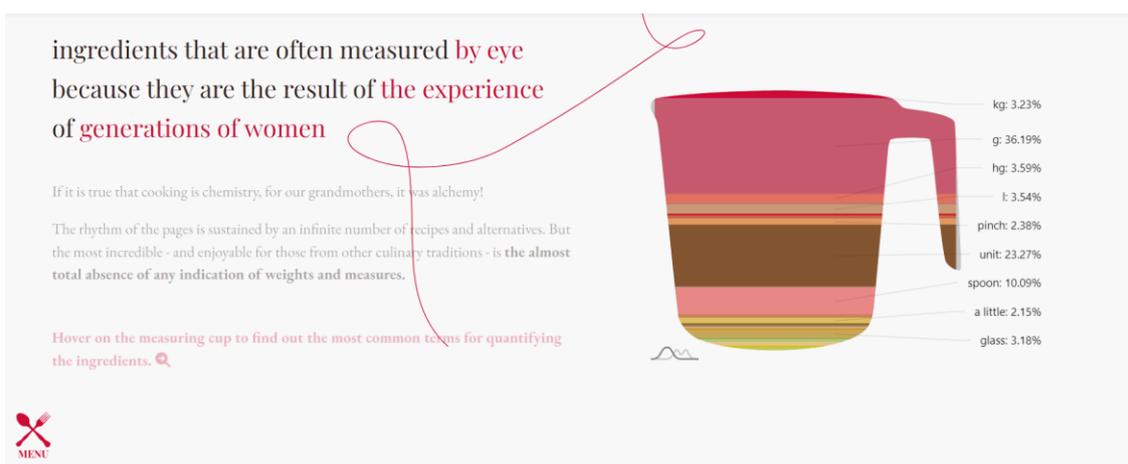

**Figure 2. An example of a chart with introductory text.**

This exploratory journey provides an overall understanding of the collection, informing users of what type of information they could expect later in the exploration of single recipes in the *recipes* section. It is also an engaging way to showcase the collection, the work in progress, and to attract new donors to enrich it continuously.

The second section of the website is the *recipes* where the user can select recipes either from an alphabetical index (like in an encyclopaedia of recipes) or can filter them by category (type of meal, ingredients, city of origin) (Figure 3). For each recipe the user can access the digitised picture of the original cookbook and the transcribed metadata (Figure 4). If we take as an example the "Pasticcio di maccheroni" recipe we discover two recipes, one by Mrs Dina and the other by Mrs Anna Maria (Figure 3). Opening both allows one to appreciate the differences in the making. This section is particularly useful to discover the variants of the names that these women used to call the ingredients, which were the most commonly used and where. Being free in the exploration allows the user to find complex recipes such as "Tortellini" that come in at least 5 different variations, or navigate into poorer recipes, perhaps dating back to war times.

## 5. CONCLUSION

Despite limited resources, the approach applied to *Ragù*, combining crowdsourcing, good practices in data management, and automated updates through a popular platform such as GitHub, showcases a pragmatic and scalable solution that can be implemented in similar projects. The digital collection's website, structured as an exploratory journey with digital storytelling elements, successfully balances overview and detailed exploration. By leveraging data visualisation techniques, it provides users with insights into the history, provenance, ingredients, dishes, and gender aspects of the cookbooks. The emphasis on an exploratory journey serves not only to enrich understanding but also to invite new contributions, ensuring a dynamic and continually expanding digital collection. Future works will address current work in progress operations, such as the inclusion of new recipes, the finalisation of the ebook-like browser of recipes, and the generalisation of the source code used to build the repository, so that similar projects can leverage out-of-the-box solutions to publish their crowdsourced collections.

Figure 3. *Recipes* search interface.

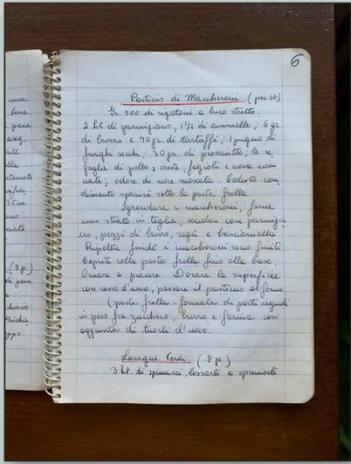

Figure 4. An example of recipe.

## 6. ACKNOWLEDGEMENTS

The *Ragù* project is directed by Dr. Mila Fumini and coordinated by the Digital Humanities Advanced Research Centre (/DH.arc, Unibo): Marilena Daquino (supervision), Giulia Manganelli (web design, web development, copywriting), and Giulia Renda (data management). Data editing: Roberta Balduzzi.
This work was partially funded by Project PE 0000020 CHANGES - CUP J33C22002850006, NRP Mission 4 Component 2 Investment 1.3, funded by the European Union - NextGenerationEU.